\documentclass{PoS}
\usepackage{graphicx,wrapfig,epstopdf}

\title{Kaon semileptonic decays near the physical point}

\ShortTitle{Kaon semileptonic decays}

\author{\speaker{K. Sivalingam}\\%
  School of Physics \& Astronomy, University of Edinburgh, EH9 3JZ, UK\\
  E-mail: \email{K.Sivalingam@sms.ed.ac.uk}}

 \author{P.A. Boyle\\
   School of Physics \& Astronomy, University of Edinburgh, EH9 3JZ, UK\\
   E-mail: \email{paboyle@ph.ed.ac.uk}}

 \author{J.M. Flynn\\
   School of Physics \& Astronomy, University of Southampton, SO17 1BJ, UK \\
   E-mail: \email{j.m.flynn@soton.ac.uk}}

 \author{A. J\"uttner\\
   School of Physics \& Astronomy, University of Southampton, SO17 1BJ, UK \\
   E-mail: \email{a.juttner@soton.ac.uk}}

 \author{C.T. Sachrajda\\
   School of Physics \& Astronomy, University of Southampton, SO17 1BJ, UK \\
   E-mail: \email{cts@phys.soton.ac.uk}}

 \author{J.M. Zanotti\\
   CSSM, School of Chemistry and Physics, The University of Adelaide,
   SA 5005, Australia \\
   E-mail: \email{james.zanotti@adelaide.edu.au}}

\author{RBC and UKQCD Collaborations}

\abstract{The CKM matrix element $|V_{us}|$ can be extracted from
  the experimental measurement of semileptonic $K\to\pi$ decays.  
  The determination depends on theory input for the corresponding
  vector form factor in QCD. We
  present a preliminary update on our  
  efforts to compute it in $N_f=2+1$ lattice QCD using domain wall fermions for
  several lattice spacings and with a
  lightest pion mass of about $170\,\mathrm{MeV}$. By using partially twisted boundary
  conditions we avoid systematic errors associated with an interpolation of 
  the form factor in momentum-transfer,
  while simulated pion masses near the physical point 
  reduce the systematic error due to the chiral
  extra\-polation.}

\FullConference{The 30th International Symposium on Lattice Field
  Theory,  June 24 - 29,  2012, Cairns, Australia}

\begin{document}

\section{Introduction}
In the Standard Model (SM) the unitary Cabibbo-Kobayashi-Maskawa (CKM)
matrix
contains information on the strength of flavour-changing weak decays as well as 
information on
correlations between different such processes.
Inconsistencies in the CKM-picture would indicate the presence of new 
physics beyond the SM. 
One therefore tries to determine all CKM-matrix elements as precisely as
possible by studying flavour changing processes both experimentally 
(e.g. at NA62 and LHCb at CERN) and theoretically.
Here we concentrate on the determination of the matrix element $|V_{us}|$ from
the study of semileptonic kaon ($K_{l3}$) decays which enables a test of
first-row CKM-unitarity, $|V_{ud}|^2 + |V_{us}|^2 + |V_{ub}|^2 = 1$.
$|V_{ub}|$ is neglected because it is smaller than the errors in the other two
terms on the l.h.s., $|V_{us}|$ and $|V_{ud}|$. 
$|V_{ud}|$ is known very precisely from neutron $\beta$-decay and
matching the corresponding precision for $|V_{us}|$ 
is crucial in searching for deviations
from CKM-unitarity and for possible signs of new
physics.

In this talk we present an update on our precision-study of
the $K_{l3}$ form factor, $f_+(0)$ \cite{Boyle:2007qe,Boyle:2010bh}.
The novelties are simulations at lighter quark masses corresponding to 
a lightest pion mass of about $170\,\mathrm{MeV}$~\cite{:2012yc} and two additional lattice 
spacings.  Both allow for  considerably improved control over systematic effects.

\section{Kaon semileptonic decays}
To date, one of the most precise determinations of $|V_{us}|$ comes from $K
\to \pi l\nu$ semileptonic decays (cf.~\cite{Colangelo:2010et}): 
The product $|V_{us}|^{2}|f_{+}(0)|^{2}$ can be determined with a precision at the
per mil level from the experimental decay rate~\cite{Antonelli:2010yf} 
and lattice QCD provides the value for $f_{+}(0)$.
Currently the level
of precision set by experiment sets the precision goal for
lattice simulations.

The form factor $f_{+}(0)$ is defined from the vector part of the
strangeness-changing weak current ($V_{\mu}=\bar{s}\gamma_{\mu}u$) according to
\begin{equation}
\label{eq:formfactor_rel}
\langle \pi(p^\prime) \big | V_\mu \big | K(p)\rangle = (p_\mu + p_\mu^\prime) f_+(q^2) + 
(p_\mu - p_\mu^\prime) f_-(q^2) \,,
\end{equation}
where $q^2=(p-p^\prime)^2$ is the momentum transfer. 
We define the scalar form factor 
\begin{equation}
 f_0(q^2)=f_+(q^2)+\frac{q^2}{m_K^2-m_{\pi}^2}f_-(q^2)\,,
\end{equation}
with $f_0(0)=f_+(0)$.  
In the SU(3) flavour limit where $m_K^2=m_\pi^2$, $f_{+}(0)\nolinebreak=\nolinebreak 1$ by
vector current conservation.
%
It can be expanded in terms of the meson masses as
%
$f_+(0) = 1 + f_2 + \ldots$
%
where $f_n=O(m_{\pi}^{n},m_K^n,m_{\eta}^n)$~\cite{Gasser:1984ux}.
$f_2$ is a known function of meson masses 
and the SU(3) pseudoscalar decay constant. 
%
%
In lattice simulations therefore effectively only the small higher order correction
%
$\Delta f = f_+(0) - (1 + f_2)$
is computed and extrapolated to the physical point.
%
%
%

%

We compute the matrix element in~(\ref{eq:formfactor_rel}) in terms of 
the ground-state contribution to suitable ratios of Euclidean two- and 
three-point functions. In a finite lattice box with periodic
boundary conditions for the quark fields, the matrix element can in this way
only be computed for meson momenta corresponding to the Fourier modes,
i.e. $\frac{2\pi}{L}\vec n$ with $n_i=\pm 0,\pm 1,\dots$ for 
a spatial volume $V=L^3$.  The form factor
at $q^2=0$ is then computed by interpolating between the data for the
form factor at these Fourier-points~\cite{Boyle:2007qe} thereby introducing
a dependence of the final result on the interpolation model;
different ans\"atze may lead to different results for
the  form factor, introducing systematic uncertanities.

\section{Partially twisted boundary condition}
%
%

%
%
%
In contrast to periodic boundary conditions, twisted boundary conditions 
allow access to hadron momenta other than $\frac {2\pi}L\vec n$~\cite{Bedaque:2004ax}.
Here we employ partially twisted boundary 
conditions~\cite{Sachrajda:2004mi} by
applying the twist only to the valence quarks, 
\begin{equation}
 q(x_{i}+L)=e^{i\,\theta_{i}} q(x_{i})\,,
\end{equation}
($\theta_{i}$ is the twist angle in the $\hat i$-direction). 
In our simulation a charged
meson of mass $m$ with one of the valence quarks twisted with 
angle $\vec \theta$
then obeys the dispersion relation
\begin{equation}
\label{eqn:pi_disprel}
E=\sqrt{m^{2}+\left(\frac{2\pi}L\vec n+{\vec{\theta}}/{L}\right)^{2}},
\end{equation}
up to exponentially supressed finite volume corrections~\cite{Sachrajda:2004mi}.
%
For matrix elements like $K \to \pi$,
where the initial and final state mesons carry twists 
$\vec\theta_i$ and $\vec\theta_f$, respectively,
the momentum transfer can be written as~\cite{Boyle:2007wg}
\begin{equation}
 q^{2}=(p_{i}-p_{f})^{2}=\Big[E_{i}(\vec{p}_{i})-E_{f}(\vec{p}_{f})\Big]^{2}-
\left[\left(\frac{2\pi}L\vec n_i +\frac{\vec{\theta_{i}}}{L}\right)-\left(\frac{2\pi}L\vec n_f +\frac{\vec{\theta_{f}}}{L}\right)\right]^{2}\,.
\end{equation}
By adjusting the twists on the initial and final meson (here $K,\pi$), 
we can evaluate
$f_{+}$ directly at $q^2=0$.
The commonly used implementation is to contruct ratios of correlation
functions, $R_{K\pi}(\vec{p}_K,\vec{p}_\pi)$ (see e.g.~\cite{Boyle:2007wg}), 
where either the kaon or
pion is kept at rest and where the twist on the other meson is tuned 
in order to obtain results~\cite{Boyle:2007wg}:
%
%
\begin{equation}\label{eq:twists}
 \begin{array}{llcccc}
&  R_{K\pi}(\vec{p}_K,\vec{0})&\textrm{with}&
  |\vec{\theta}_K| =
           L\sqrt{\left({m_K^2+m_\pi^2 \over 2m_\pi}\right)^2 - m_K^2}
      &\textrm{and}&\vec{\theta}_\pi=\vec{0}\,,\\
{\rm and}&  R_{K\pi}(\vec{0},\vec{p}_\pi)&\textrm{with}&
          |\vec{\theta}_\pi| =L
          \sqrt{\left({m_K^2+m_\pi^2 \over 2m_K}\right)^2 -
          m_\pi^2}&\textrm{and}
      &\vec{\theta}_K =\vec{0}\,.\\
 \end{array}
\end{equation}
The form factor  $f_0(0)$ can then be evaluated directly at
$q^2=0$ as 
\begin{equation}\label{eq:lin_comb}
f_0(0)=\frac{ R_{K\pi}(\vec{p}_K,\vec{0})(m_K-E_\pi)
- R_{K\pi}(\vec{0},\vec{p}_\pi)(E_K-m_\pi) }{
(E_K+m_\pi)(m_K-E_\pi)-(m_K+E_\pi)(E_K-m_\pi)
}\,.
\end{equation}
This expression is derived by considering only the time-component 
of the vector current matrix element (\ref{eq:formfactor_rel}).
Using all the other components of the weak vector current, we can
construct an over-constrained system of equations
\begin{eqnarray}\label{eq:simul_eqn}
R_{K\pi}(\vec{\theta}_K,\vec{0},V_t) &=& 
    f_+(0)\,(E_K+m_\pi) + f_-(0)\,(E_K-m_\pi)\,,\nonumber\\
R_{K\pi}(\vec{0},\vec{\theta}_\pi,V_t) &=& 
    f_+(0)\,(m_K+E_\pi) + f_-(0)\,(m_K-E_\pi)\,,\nonumber\\
R_{K\pi}(\vec{\theta}_K,\vec{0},V_i) &=& 
    f_+(0)\,\theta_{K,i} + f_-(0)\,\theta_{K,i}\,,\nonumber\\
R_{K\pi}(\vec{0},\vec{\theta}_\pi,V_i) &=& 
    f_+(0)\,\theta_{\pi,i} -
    f_-(0)\,\theta_{\pi,i}\ ,\qquad (i=x,y,z)\,,
\end{eqnarray}
which we solve to obtain the form factors directly at $q^2=0$. Note that 
distributing the twist over all spatial directions maximises the number of 
non-trivial equations. Having
eliminated the systematic error due to the $q^2$ interpolation,
the remaining dominant source of systematic uncertainty is due to
the chiral extrapolation of the lattice data to the physical point
\cite{Boyle:2010bh}.
%

\section{Results}
In~\cite{Boyle:2007qe} we presented results for the form factor 
from simulations of domain wall fermions (DWF) with 
the Iwasaki Gauge action,  an inverse lattice spacing of 1.7GeV and 
$m_\pi$ in the range $330$--$700\,\mathrm{MeV}$. 
Here we extend the analysis by adding  new ensembles 
32Fine~\cite{Aoki:2010dy}, 32Coarse~\cite{Arthur:2012yc}
(cf. table~\ref{tab-ensproperties}) providing data for considerably lighter
pion masses down to $170\,\mathrm{MeV}$ and two additional lattice spacings.
\begin{table}
\vspace{-0.4cm}
\centering
\small{
\begin{tabular}{c|c|c|c|c}
\hline\hline
\rule{0cm}{0.4cm}Label & Size & $S_G$ &  $a^{-1}$ & $m_\pi(\,\mathrm{MeV})$  \\
\hline
\rule{0cm}{0.4cm}24Coarse & $24^3\times 64\times 16$ & Iwasaki 		& 1.75(4)  & $330$, $420$, $550$ , $670$ \\
\rule{0cm}{0.4cm}32Fine &$32^3\times 64\times 16$ & Iwasaki 		& 2.31(4)  & $290$, $350$, $400$ \\
\rule{0cm}{0.4cm}32Coarse & $32^3\times 64\times 32$ & Iwasaki+DSDR 	& 1.37(1)  & $170$, $250$\\
\end{tabular}
}
\caption{A summary of the three ensembles used in this analysis. Here
  `$S_G$' denotes the Gauge action, `$m_\pi$' the pion mass and
  $a^{-1}$ the lattice spacing.  The 32Fine and 32Coarse data are new
  in this calculation.}
\label{tab-ensproperties}
\end{table}
\begin{figure}[tbp]
\begin{minipage}{0.5\linewidth}
\centering
\includegraphics[scale=0.39]{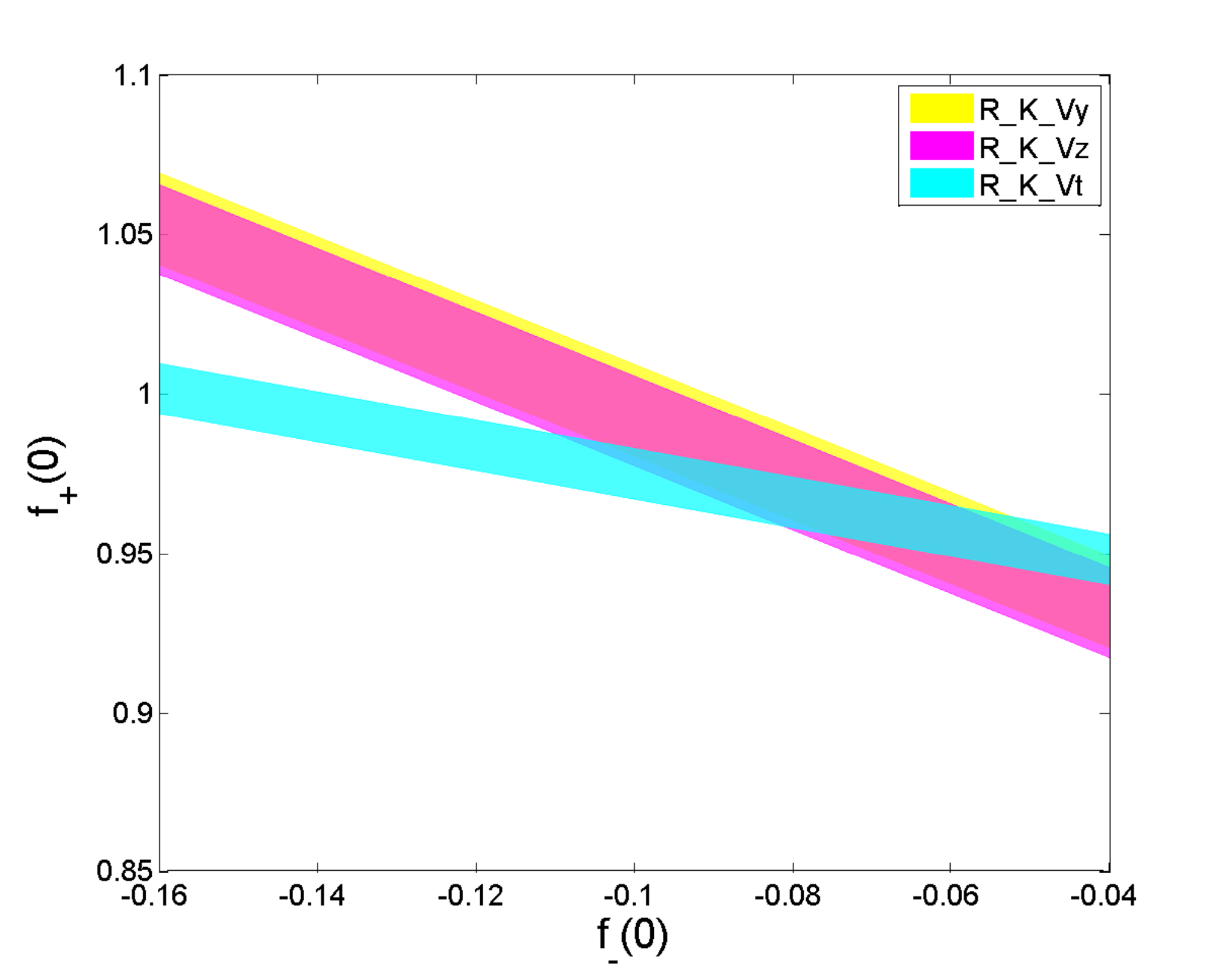}
\end{minipage}%
\begin{minipage}{0.5\linewidth}
\centering
\includegraphics[scale=0.39]{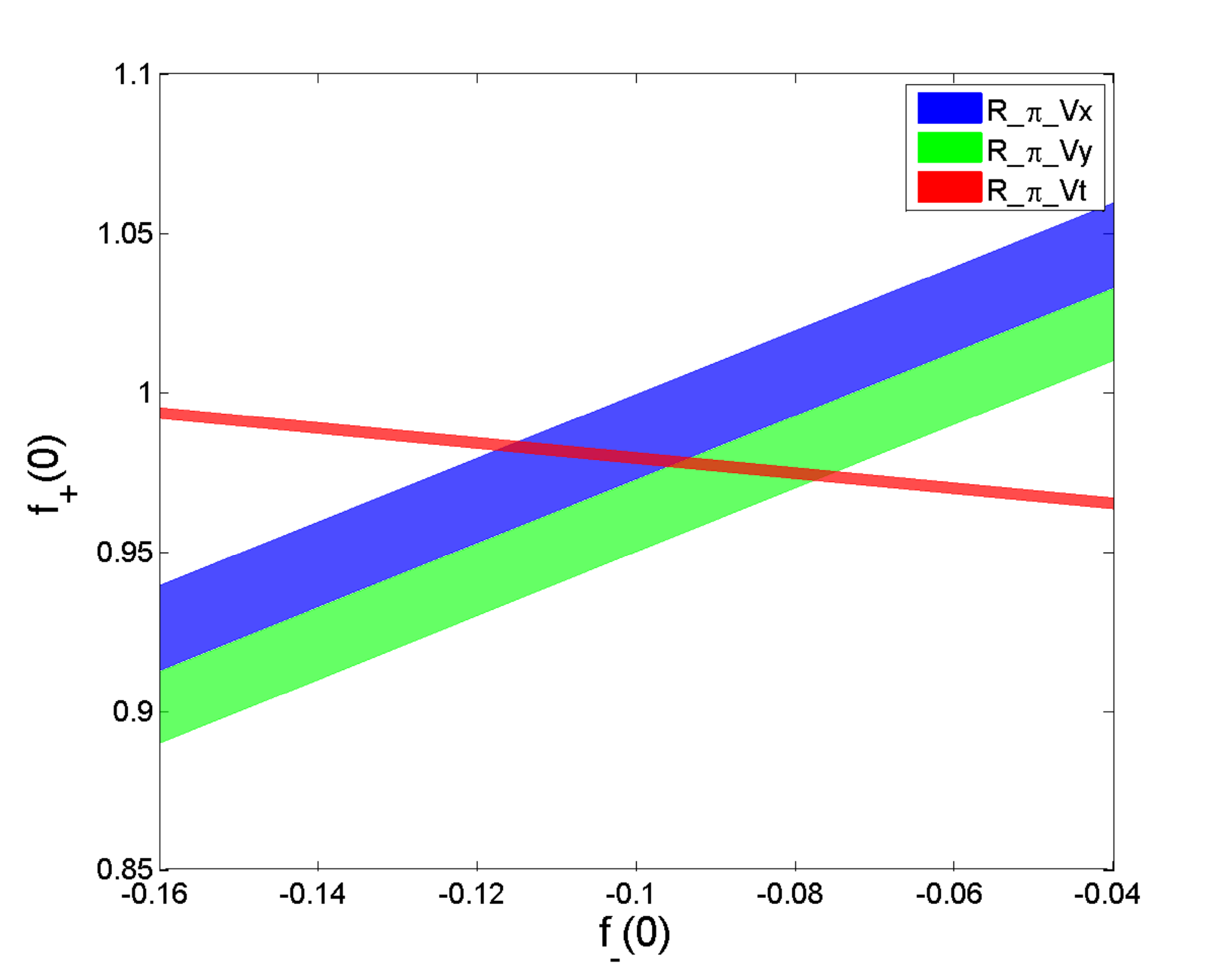}
\end{minipage}
\caption{Plot of $f_+(0)$ vs. $f_-(0)$ for kaon(left) and pion(right)
only twisted equations}
\label{fig-pi_ka_twist}
\end{figure}

Closer to the physical point the twist-angles prescribed
by eqs.~(\ref{eq:twists}) are very large, in particular for the case where
the kaon is moving and the pion is at rest. This manifests itself in larger
statistical fluctuations for the ratio $R_{K \pi}(\vec p_K,\vec 0)$.
%
In order to find a better choice of kinematics we analysed the situation 
further:
%
From eq.~(\ref{eq:simul_eqn}), the slope of $f_{+}(0)$ with respect to 
$f_{-}(0)$ is given by
\begin{equation}
  \frac{\partial f_{+}(0)}{\partial f_{-}(0)}\bigg|_{\theta_{K}=0} =
-\frac{m_{K}-E_{\pi}}{m_{K}+E_{\pi}}\,,\qquad
  \frac{\partial  f_{+}(0)}{\partial f_{-}(0)}\bigg|_{\theta_{\pi}=0} = 
-\frac{E_{K}-m_{\pi}}{E_{K}+m_{\pi}}\,,
\end{equation}
for the matrix element of the time-component of the vector current, $V_t$, and
\begin{equation}
  \frac{\partial f_{+}(0)}{\partial f_{-}(0)}\bigg|_{\theta_{K}=0} = 1
  \qquad 
  \frac{\partial f_{+}(0)}{\partial f_{-}(0)}\bigg|_{\theta_{\pi}=0}=-1\,,
\end{equation}
for the spatial components $V_{x,y,z}$.
\begin{figure}[tbp]
\begin{minipage}{0.5\linewidth}
\centering
    \includegraphics[scale=0.39]{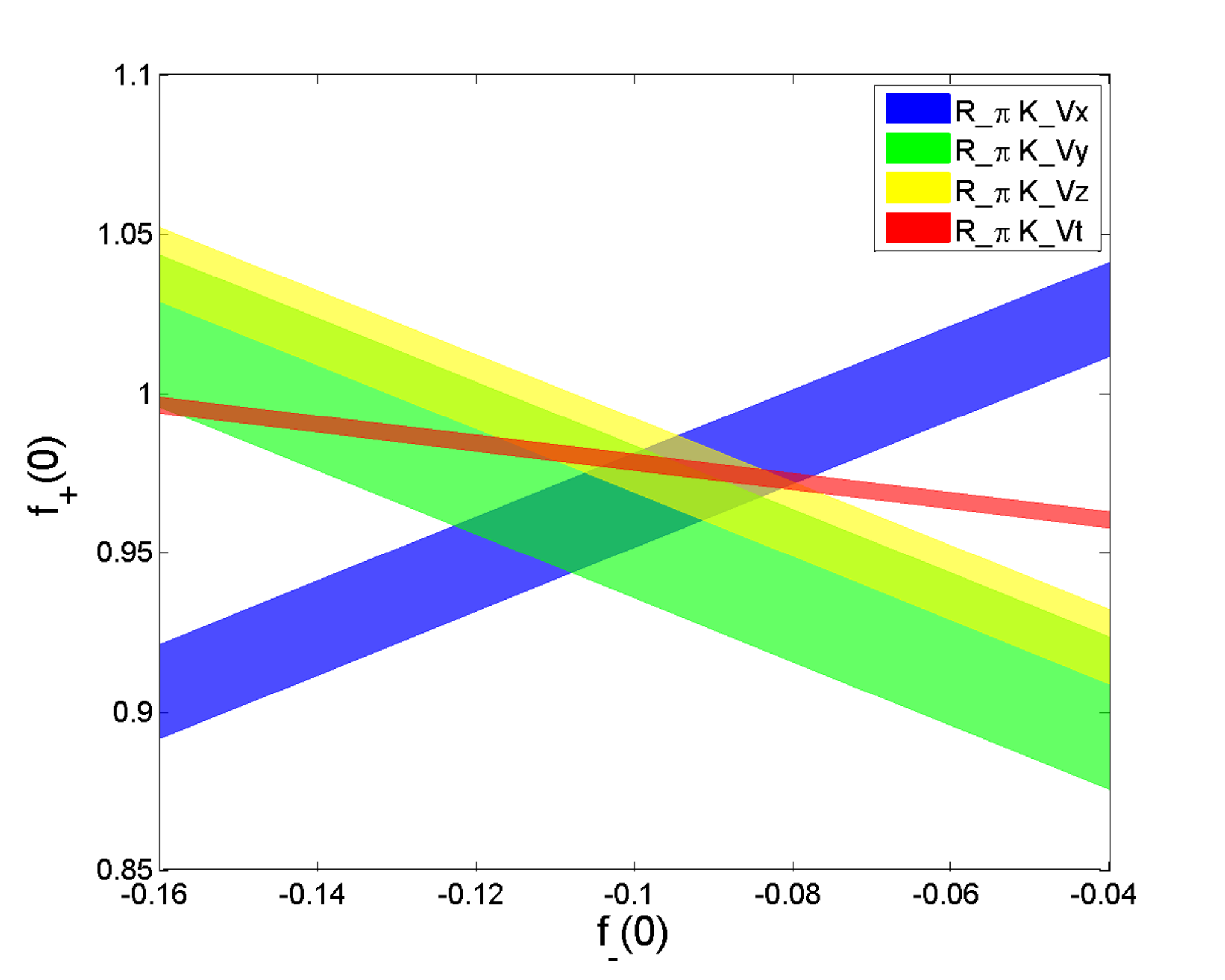}
\end{minipage}%
\begin{minipage}{0.5\linewidth}
\centering
    \includegraphics[scale=0.39]{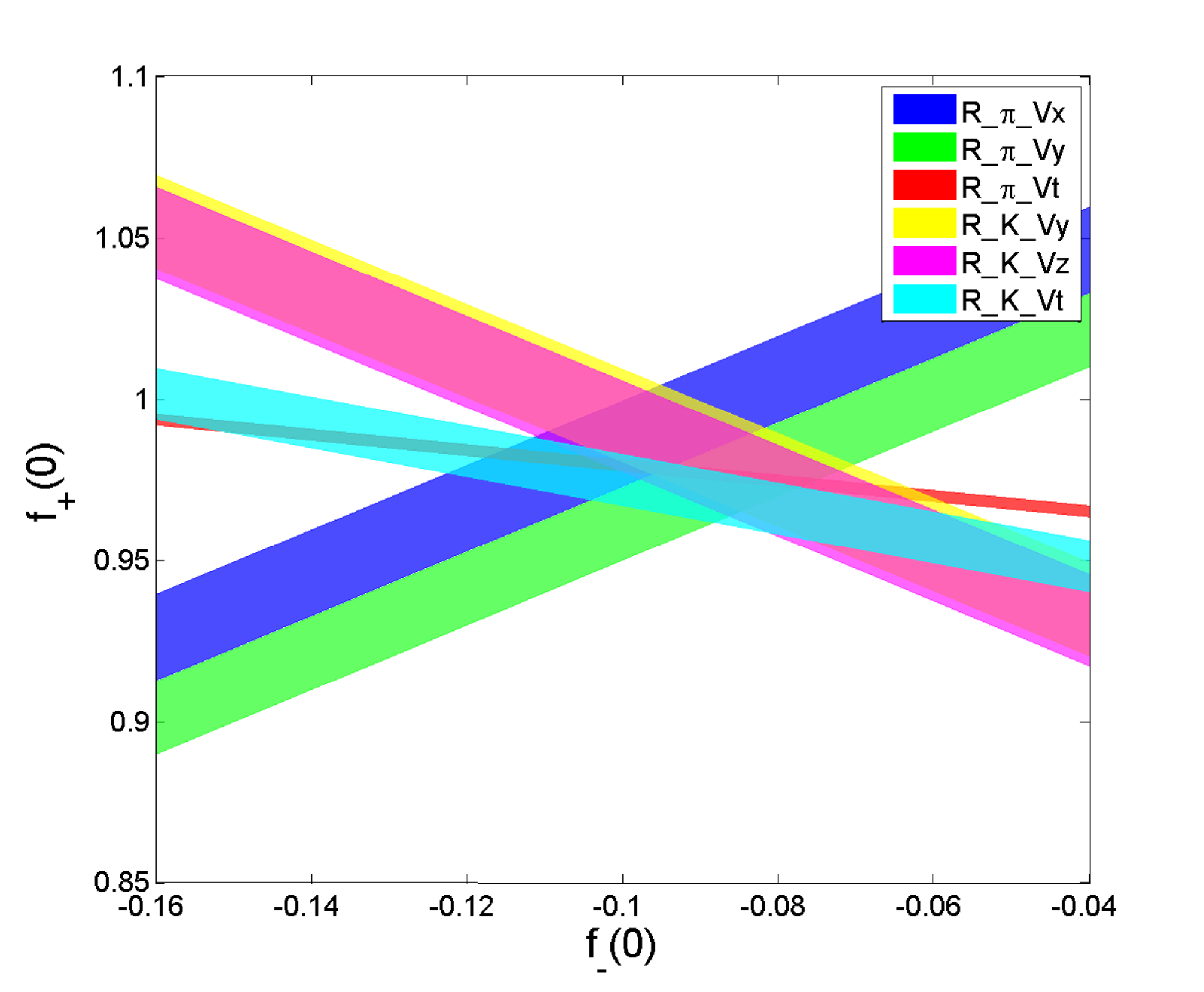}
\end{minipage}
  \caption{Plot of $f_+(0)$ vs. $f_-(0)$ for the cases where (left)
    both pion and kaon are twisted (right) pion only and kaon only
    twisted equations are combined.}
\label{fig-pi_and_ka_twist}
\end{figure}
The solutions for all equations for ($\vec p_K,\vec 0$) and 
($\vec 0,\vec p_\pi$) on the 32Coarse ensemble with $m_\pi=250\,\mathrm{MeV}$
are shown in the l.h.s. and r.h.s. plots in 
Fig~\ref{fig-pi_ka_twist}, respectively.
While all solutions 
have a negative slope for the case where only the kaon is twisted, there
are solutions with opposite slopes in the case where the pion is twisted.
%
Because the solution is
given by the intersection of the individual constraints, the
kinematical situation where the pion is moving (twisted) and the
kaon is at rest provides the best result. The statistical errors
are also smaller in this case.
Motivated by these observations, we computed all correlation functions once
again for a third choice of kinematics with  $q^2=0$,
where both the kaon and the pion are twisted. 
This leads to a good constraint for $f_+(0)$ as shown in the left plot in 
Fig.~\ref{fig-pi_and_ka_twist}. 
The result is in agreement with the result 
obtained by solving all simultaneous equations for the cases
where either the pion or the kaon are twisted as shown in the right plot of Fig.~\ref{fig-pi_and_ka_twist}
(obtained by combining the plots in Fig.~\ref{fig-pi_ka_twist}).

We now turn our attention to the mass- and momentum-dependence of the results.
In principle, using partially twisted boundary conditions, one is independent of
the momentum interpolation. We wish however to include our earlier data
sets, the three heavier ensembles in the 24Coarse set, for which we have not
generated data directly at $q^2=0$. From~\cite{Boyle:2007qe} we know 
that in these cases the interpolation introduces hardly any model-dependence.
In addition to the momentum-dependence a fit ansatz should also incorporate 
the $\mathrm{SU}(3)$-symmetry-breaking nature of $f_+(0)$ and the strange 
quark mass dependence since the simulated strange quark mass is
not exactly at the physical point. Our ansatz
%
%
with four fit parameters
$A_0,A_1,M_0,M_1$ \cite{Boyle:2007qe} is
\begin{equation}
f_0(q^2,m_\pi^2,m_K^2) =
\frac{1+f_2+(m_K^2-m_\pi^2)^2(A_0 + A_1(m_K^2+m_\pi^2))}
{1-q^2/(M_0+M_1(m_K^2+m_\pi^2))^2}\,.
\label{eq:global}
\end{equation}
\begin{figure}[htbp]
\begin{center}$
\begin{array}{cc}
  \includegraphics[scale=0.42]{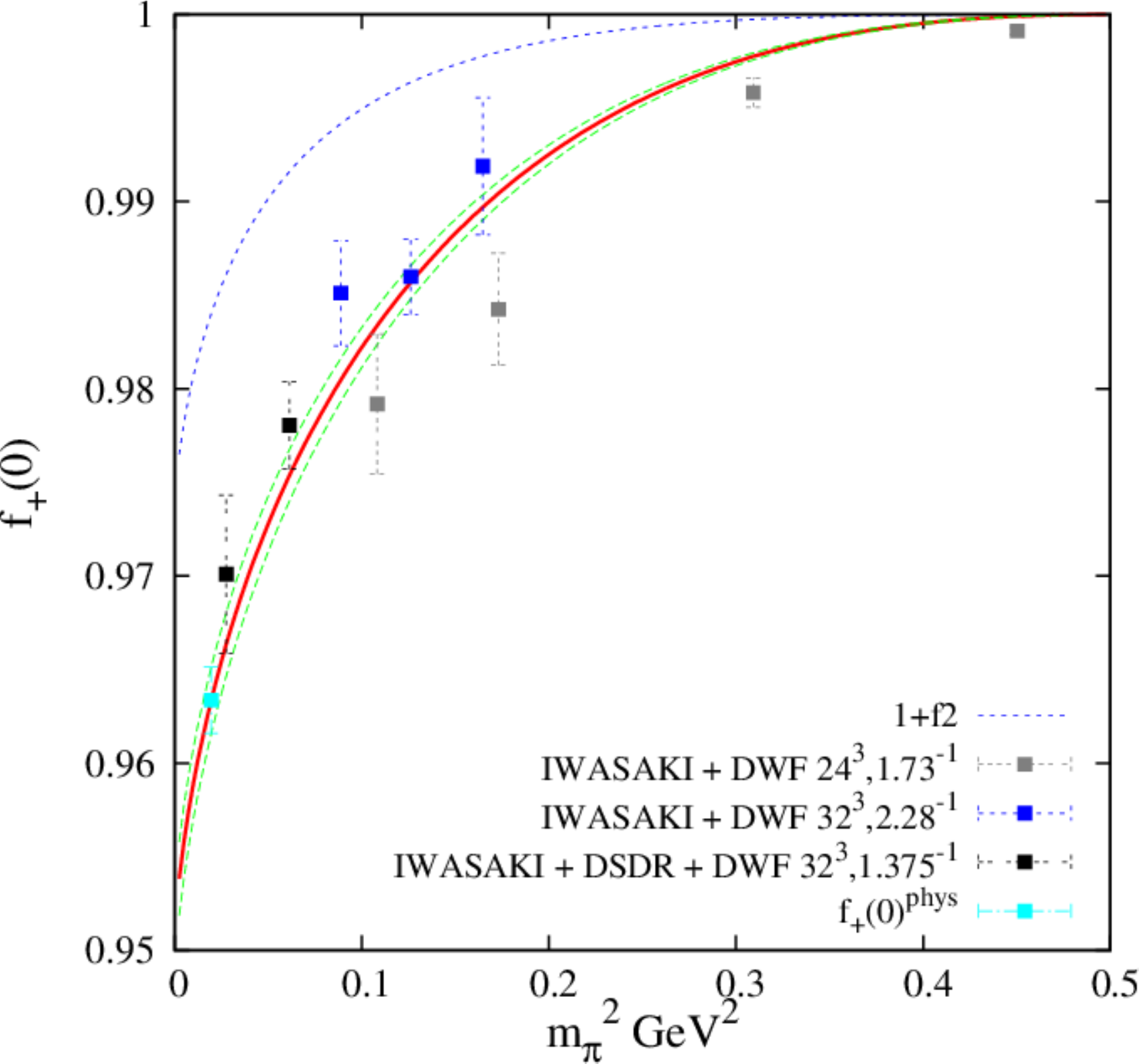} &
  \includegraphics[scale=0.42]{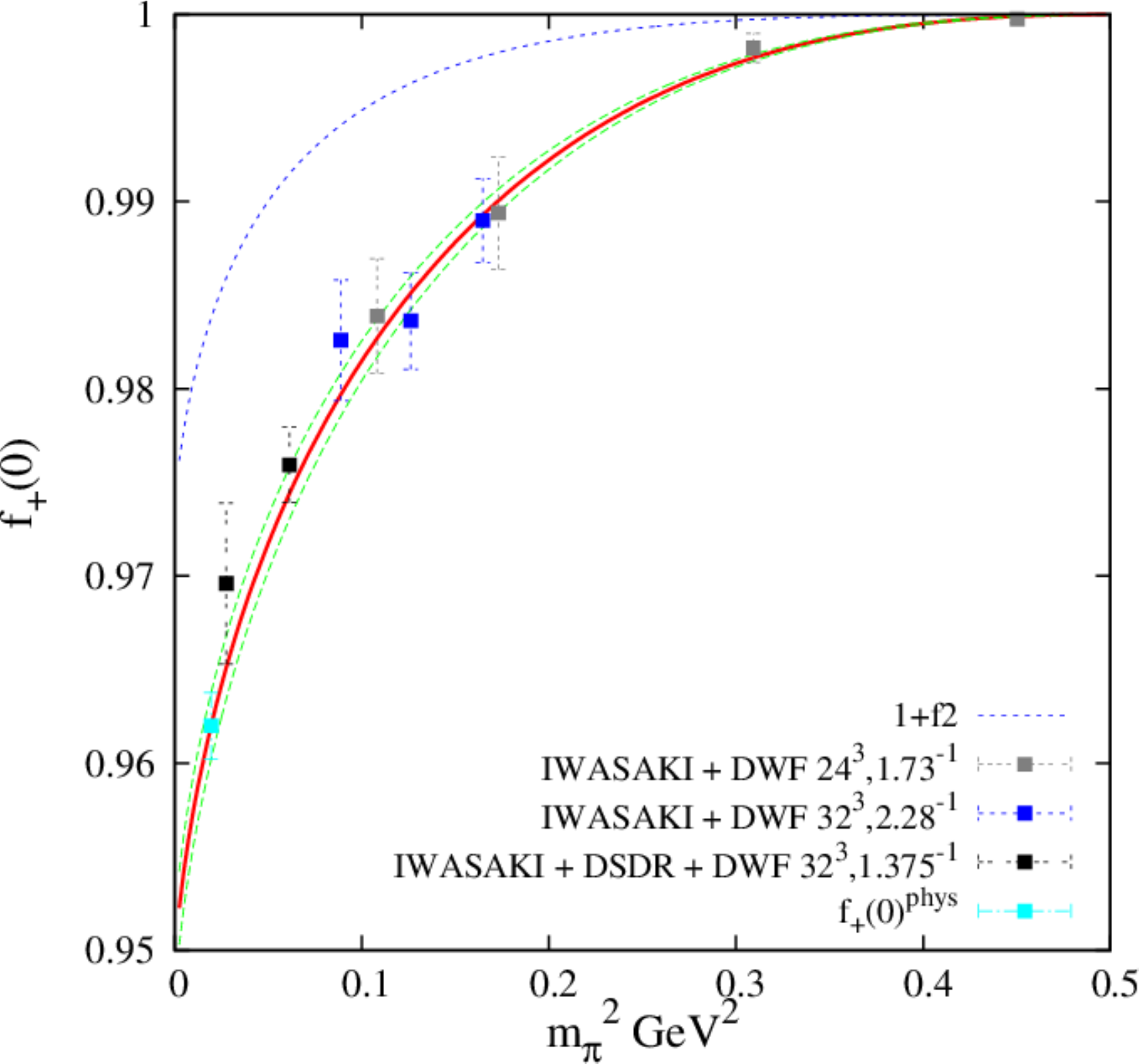}
\end{array}$
\end{center}
\caption {Plot of $f_+(0)$ dependence on $m_\pi^2$. 
Left : raw data points at the simulated (uncorrected) strange quark
masses.  
Right : The data points are shifted to physical strange quark mass.
In both plots, the curve uses a parameterisation for the kaon mass
with strange quark held fixed at its physical value.}
\label{fig-f0}
\end{figure}
Fig~\ref{fig-f0} (left) summarizes the results of the analysis for
all ensembles.
The data points are the ones for the simulated, i.e. unphysical strange-quark
mass. 
After correcting towards the physical strange-quark mass using the ansatz in 
eq.~(\ref{eq:global}), all data points line up on the fit-curve
in the r.h.s. plot in Fig.~\ref{fig-f0}.
%
%
Our preliminary result for $f_+(0)$ at the physical point is
indicated by the light blue square. 
At this early stage of the analysis we find that the statistical error at 
the physical point has been reduced with respect to our earlier 
results~\cite{Boyle:2010bh}.

In our fit ansatz we use the decay constant in the chiral
limit, $f_0$, in the NLO term and have a form for the NNLO term which
together make the ansatz consistent with the Ademollo-Gatto theorem
and symmetry under interchange of $m_\pi$ and $m_K$. Since we do not
know the precise value for $f_0$, we repeat the global fit for $f_0 =
100, 115$ and $131\,\mathrm{MeV}$. This variation in $f_0$ is also
assumed to account partially for our NNLO term not being the full
chiral perturbation theory expression. The result for $f_+(0)$ using
$f_0 = 115\,\mathrm{MeV}$ is taken as our best value and the values at
$100$ and $131\,\mathrm{MeV}$ provide an estimate of the systematic
error from the chiral extrapolation.

\begin{figure}[htbp]
  \begin{center}
    \includegraphics[scale=0.4]{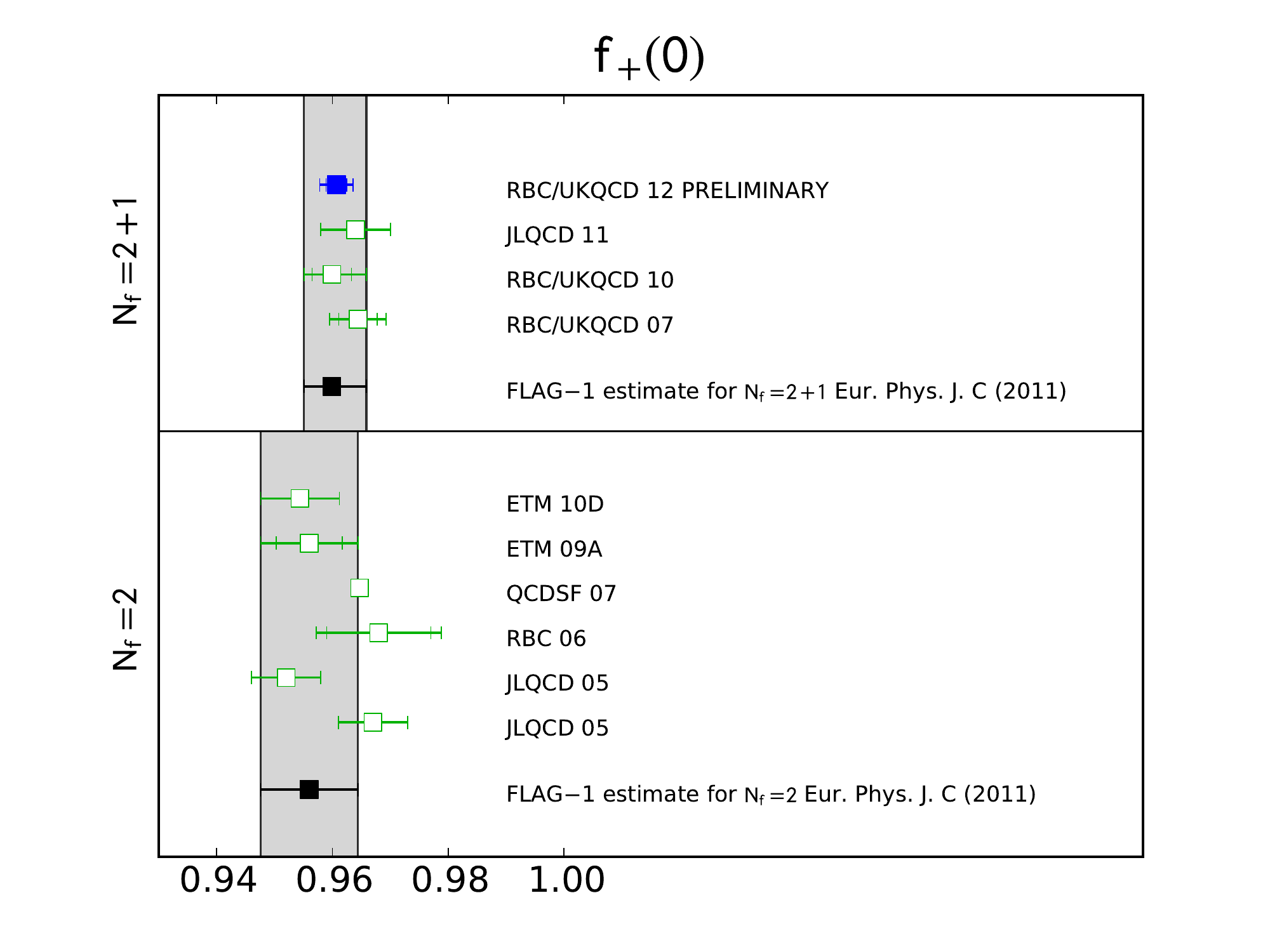}
  \end{center}
  \caption{ Comparison of recent Lattice results}
\label{fig-recentresults}
\end{figure}

In Fig.~\ref{fig-recentresults} we compare our new preliminary result 
for $f_+(0)$ with other recent Lattice determinations.
We emphasize that our result is preliminary, but we expect 
to reach a new level of precision when the analysis is complete.

\section{Conclusion}

We have presented an update on our  simulations aiming at the precise
prediction of the kaon semileptonic form factor at vanishing momentum
transfer. The novelties are simulations for pion masses down to $170\,\mathrm{MeV}$ and
for  three different  lattice spacings.
We identified preferred choices for the kinematics when using partially
twisted boundary conditions in order to simulate directly at $q^2=0$.
In particular, we now understand why 
twisting either only the pion or both the kaon and the pion 
can constrain the form factor better than  twisting only the kaon.
%
%
%
At this early stage of the analysis the inclusion of the new ensembles in 
our global fit leads to a reduction of the statistical error at the
physcial point compared to our earlier result~\cite{Boyle:2010bh}.
This indicates significant progress in the study of K $\to$ $\pi$
semileptonic form factors which would directly impact our knowledge of the
CKM Matrix and provide for improved constraints for new physics beyond
the Standard Model.
%

\section{Acknowledgement}

We thank our colleagues in RBC and UKQCD within whose programme this
calculation was performed. 
The computations were done using the STFC's DiRAC facility at Swansea,
JUGENE at the J\"ulich Supercomputing Centre and the DiRAC
facility at Edinburgh.
KS acknowledges support by the European Union under the Grant
Agreement number 238353 (ITN STRONGnet);
CTS and JMF by STFC grant ST/J000396/1 and ST/H008888/1;
P.A.B, by STFC grants ST/K000411/1, ST/J000329/1 and ST/H008845/1;
AJ by  European Research Council (FP7/2007-2013)/Grant agreement 27975;
and JMZ by the Australian Research Council under grant FT100100005.

\end{document}